\documentclass[12pt]{iopart}
\usepackage{graphicx,iopams}
\begin{document}
\title[Vortex pseudomomentum and dissipation]
{Vortex pseudomomentum and dissipation in a superfluid vortex lattice}
\author{H M Cataldo}
\address{Departamento de F\'{\i}sica, Facultad de Ciencias Exactas y
Naturales, Universidad de Buenos Aires
 and 
Consejo Nacional de Investigaciones Cient\'{\i}ficas y T\'ecnicas,
Argentina}
\ead{cataldo@df.uba.ar}
\begin{abstract}
We propose an alternative approach to the dissipative vortex dynamics 
occurring in a superfluid vortex lattice at finite temperatures.  
Focusing upon the pseudo\-momentum of a vortex and its surrounding quasi\-particles, we derive 
an equation of
motion which, in spite of yielding the same evolution as the usual one for massless vortices, 
does not involve the vortex mass.
This picture could provide further insights into the
 controversy about the nature of the vortex mass. 
\end{abstract}

\maketitle

\section{Introduction}\label{s1}
Below the lambda transition, at temperatures below 
$T_\lambda=2.172$ K, liquid helium is known as {\it helium II} and
can be regarded as a mixture
of a normal fluid with mass density $\rho_n$ and a superfluid with mass density $\rho_s$.
Such densities are temperature dependent, so that $\rho_s(T_\lambda)=\rho_n(0)=0$,
while the total density $\rho=\rho_s+\rho_n$ remains nearly constant. The most striking 
property of such a superfluid component is, perhaps, that it can only rotate through vortices
having microscopic cores and quantized circulations. In fact, the
circulation of the superfluid velocity field around each of such vortices is quantized in
units of $h/m_4$, the so-called quantum of circulation $\kappa$, given by the ratio of 
Planck's constant and the mass of one $^4$He atom. In practice, however, we only
find configurations of one quantum per vortex, since they are favored by energetic 
considerations \cite{don}. When a rotating sample of liquid helium is cooled below the lambda
temperature, all the rotation of the superfluid becomes concentrated in such vortices, which
eventually form a lattice consisting in a
uniform array of vortex filaments parallel to the axis of rotation \cite{don,bar,sonin}.
However, the {\em macroscopic} superfluid velocity field, which corresponds to spatial averages
over regions large compared with the spacing between vortices,
yields the usual configuration of solid body flow,
${\bf v}_s({\bf r})=\Omega_{\rm rot}{\bf \hat z}\times{\bf r}$
for a rotation frequency $\Omega_{\rm rot}$ around the $z$ axis.
On the other hand, the {\em microscopic} superfluid velocity field, ie without averaging, is irrotational except
where a vortex is located. Since the circulation of such a field around a vortex yields the quantum
of circulation $\kappa$, according to
the Stokes' theorem we may write
\begin{equation}
\frac{1}{A}\int\int_A dx\,dy\,{\bf \hat z}\cdot{\bf w}=\kappa N_A/A,
\label{1.8}
\end{equation}
where the area $A$ in the $x$-$y$ plane contains $N_A$ vortices and
${\bf w}$ denotes the microscopic vorticity, ie the curl of the microscopic superfluid
velocity. On the left-hand side of (\ref{1.8}) we have an averaged vorticity which should be
identified with the macroscopic value ${\bf \nabla}\times{\bf v}_s=2\Omega_{\rm rot}\,
{\bf \hat z}$.
Thus, the above equation yields a link between the macroscopic and the microscopic views of
the superfluid component, namely the number of vortices per unit area corresponds to 
the ratio of the macroscopic vorticity to the quantum of circulation,
$N_A/A=2\Omega_{\rm rot}/\kappa$ (Feynman's rule \cite{don,feyn}).

Just as the superfluid flow is microscopically formed by vortices, the normal fluid consists of
superfluid quasiparticle excitations, phonons and rotons, the average flow of which is 
characterized
by a normal fluid velocity field ${\bf v}_n$ given by 
$\Omega_{\rm rot}{\bf \hat z}\times{\bf r}$. 
That is, both fluids are expected to move with the same velocity
at equilibrium. It is important to recall, however, that there
are remarkable exceptions to this result, eg the 
 equilibrium configuration of the Hess-Fairbank experiment \cite{hess},
or the {\it metastable} superflow states having effectively infinite
lifetimes \cite{leggett}. Now, keeping the 
focus on the simpler situation of having identical normal 
and superfluid velocity fields at an equilibrium state to be reached within
a finite relaxation time, one may interpret that such a behaviour arises, 
from a microscopic viewpoint, from the vortex
motion with the normal fluid velocity in order to avoid dissipation.
That is, any relative motion of vortices with
respect to the normal fluid in their vicinity, should be subjected to a friction force that
causes such a motion to eventually cease. Such a {\it mutual friction force} \cite{hall} between the two 
fluids appears then as playing a central role
in the mechanism which maintains the stability of the above equilibrium state.
A phenomenological model for this macroscopic dynamics was proposed long ago through the so-called
Hall-Vinen-Bekharevich-Khalatnikov (HVBK) equations \cite{hall,bek}.
These basically consist of a Navier-Stokes equation for the normal fluid and an Euler equation
for the superfluid, which, in the absence of pressure and temperature gradients, are coupled together
only by a mutual friction term \cite{don,bar}. The original proposal of the HVBK model 
 was later rederived from first 
principles within the framework of classical continuum mechanics \cite{hills},
but a derivation from a full microscopic theory is still lacking.
On the other hand, with the exception of a few works \cite{hend,peral}
and because of their complexity, 
the HVBK equations have been mainly utilized so far
to model helium II with a spatially uniform configuration of vortices.

The simplest departure from the rotating equilibrium configuration is given by a normal
fluid field of the form $(\Omega_{\rm rot}+w_0)\,{\bf \hat z}\times{\bf r}$,
but unfortunately a simple ansatz like 
${\bf v}_n({\bf r},t)=\Omega_n(t)\,{\bf \hat z}\times{\bf r}$ does not
constitute an acceptable solution of the 
HVBK equations. 
Simplicity, however may be preserved by changing to a different geometry consisting in
rectilinear flows of uniform vorticity,
\begin{eqnarray}
{\bf v}_s({\bf r},t)&=&-2\,\Omega_s(t)\,y\,{\bf \hat x}\label{1.1}\\
{\bf v}_n({\bf r},t)&=&-2\,\Omega_n(t)\,y\,{\bf \hat x}\label{1.2}
\end{eqnarray}
($y<0$), where $\Omega_s(t)$ and $\Omega_n(t)$ should converge for $t\rightarrow\infty$ to 
a common steady state value. Then, to make contact with the standard rotational configuration,
we may identify such a value with the former angular velocity $\Omega_{\rm rot}$.
Note that this assignment leads to a uniform vorticity ${\bf \nabla}\times{\bf v}_s
=2\,\Omega_{\rm rot}\,{\bf \hat z}$, which coincides with that of the rotational scheme.

The above macroscopic view of 
the interaction between superfluid and normal fluid,
ruled by the HVBK equations, is complemented by the microscopic picture of a dissipative
vortex dynamics arising from the scattering of quasiparticles by vortex lines\footnote{Vortex 
bending in the form of  thermal excitation of vortex oscillation modes, or collective excitations as
Tkachenko waves are not expected to be relevant to this discussion \cite{don,cond,jpa}.}. 
In the usual approach, the vortex equation of motion arises simply from assuming
a vanishing total force, which is given by the sum of a hydrodynamic Magnus force and a dissipative
force. Such a neglect of
 the vortex inertia corresponds to the assumption that its mass is given by the hydrodynamic mass of a
core of atomic dimensions \cite{don}.
This result, however, has never been experimentally confirmed owing to the difficulties that 
embodies a
direct measurement of the vortex mass \cite{snap}.
Moreover, there are different theories \cite{popov,duan,tang} that
yield several orders of magnitude higher values for the vortex mass, 
casting doubt on models based on massless vortices\footnote{There have also been conflicting results 
for the vortex mass in superconductors \cite{mass}.}.
On the other hand, it has been recently suggested that an unambiguous vortex mass may not exist,
and that inertial effects in vortex dynamics may be scenario-dependent \cite{thou}.
Given such an open debate,
it seems to be quite advisable to follow an eclectic procedure,
assuming in what follows a finite vortex
mass per unit length, which we shall denote through the parameter $m_v$.

 There is a close analogy between the dynamics 
of massive rectilinear
 vortices and the well-known electrodynamical problem of
point charges subjected to a uniform magnetic field and a perpendicular electric field.
More precisely, there exists a whole 
mapping by which a 2-D homogeneous superfluid can be mapped onto a (2+1)-D electrodynamic 
system, with vortices and phonons playing the role of charges and photons, respectively 
\cite{popov,ambe,arovas,fischer,lundh}.
Here we shall only make a restricted use of this mapping, which corresponds to
 the formal analogy between the
Magnus and Lorentz forces.
Thus we may assume  that the dissipative dynamics of the vortex lattice 
should be ruled by three characteristic frequencies, namely the imposed rotational frequency 
$\Omega_{\rm rot}$, an initial departure $w_0$ from this frequency and the cyclotron frequency stemming 
from the electromagnetic analogy \cite{jpa}
\begin{equation}
\Omega=\rho_s\kappa/m_v.\label{2}
\end{equation}
Taking as a lowest estimate of the vortex mass the value arising from the above hydrodynamical model,
we have $\Omega\lesssim  3\times 10^{12}$ s$^{-1}$ \cite{don}.
On the other hand, typical experimental values are of order $\Omega_{\rm rot}\sim 1$ s$^{-1}$,
so we shall restrict our study to cyclotron frequency values fulfilling
 $\Omega_{\rm rot}\ll \Omega$.

The above assumption of massive vortices and rectilinear flows leads us to a very
useful magnitude, which to the best of our knowledge has not been utilized so far, that is the 
concept of {\em vortex pseudomomentum}. In fact, such a pseudomomentum corresponds to the vortex
generator of  translation, as can be straightforwardly shown from the electromagnetic analogy
\cite{yosh}. More generally, the translation generated by a pseudomomentum corresponds to 
a motion of the physical state (vortex) but keeping the medium, the uniform superfluid in this case, 
fixed \cite{peie}. 
In addition, contrary to the other two momenta ({\it canonical} and  {\it dynamical}) 
that can be ascribed to a vortex, the vortex pseudomomentum turns out to be
 free from the ambiguities carried by the vortex mass.
Here it is worthwhile noticing that a similar situation occurs in the case of the normal fluid. 
In fact, it can be
shown that the momentum of a sound wave (or of a phonon in quantum mechanics), turns out to be a very
complicated object, which may not even have a well-defined value at all, while its pseudomomentum
is a simple quantity and far more useful \cite{peie,stone,b-m}. This led us to investigate a pseudomomentum
approach to the 
dissipative vortex dynamics occurring in a uniform vortex lattice, finding that
this formalism leads to the same
evolution as that predicted by the usual approach,
whereas it involves far less restrictive assumptions about the value of  
the vortex mass.

This paper is organized as follows.
In the next section we study a solution of the HVBK equations for the 
above rectilinear flows.
In section \ref{s3} we analyze the dissipative vortex 
dynamics from the viewpoint of the usual phenomenological approach.
A Hamiltonian approach is proposed in 
section \ref{s4}, which consists of a vortex Hamiltonian based on the analogy to electrodynamics
(section \ref{sec1A}) and a Hamiltonian for the quasiparticle gas representing the
normal fluid (section \ref{sec1B}). Finally a pseudomomentum equation of motion 
is obtained and discussed in section \ref{sec1C}.

\section{Two-fluid equations}\label{sec2}
The HVBK equations for the rectilinear flows (\ref{1.1})-(\ref{1.2}) are very simple and read
\begin{eqnarray}
\rho_n\frac{\partial{\bf v}_n}{\partial t} & = & {\bf F}\label{p1}\\
\rho_s\frac{\partial{\bf v}_s}{\partial t} & = & -{\bf F}\label{p2},
\end{eqnarray}
where the mutual friction force \cite{don,bar,hall},
\begin{equation}
{\bf F}=\frac{\rho_n\rho_s}{\rho}\,\Omega_s(t)
[-B({\bf v}_n-{\bf v}_s)+B'\,{\bf \hat z}\times({\bf v}_n-{\bf v}_s)],
\label{p3}
\end{equation}
becomes the single coupling term between both fluids and
the dimensionless coefficients $B$ and $B'$
weight dissipative and nondissipative contributions
to such a force.
If we assume that the system is not far from equilibrium, so that the factor $\Omega_s(t)$ in (\ref{p3})
can be approximated by $\Omega_{\rm rot}$, replacing in
(\ref{p1}) to (\ref{p3}) the velocity fields according to
(\ref{1.1}) and (\ref{1.2}), 
 an elementary calculation leads to the solution
\begin{eqnarray}
\Omega_n(t)&=&\Omega_{\rm rot}+w_0\,e^{-\Omega_{\rm rot}Bt}\label{1.6}\\
\Omega_s(t)&=&\Omega_{\rm rot}-\frac{\rho_n}{\rho}\,w_0\,e^{-\Omega_{\rm rot}Bt}\label{1.7},
\end{eqnarray}
where we should assume temperatures below 1 K, so that the normal 
fluid density turns out to be much smaller than the superfluid
one ($\rho_n/\rho_s< 10^{-3}$) and the last term in (\ref{1.7}) becomes negligible, ie
\begin{equation}
\frac{\rho_n w_0}{\rho_s\Omega_{\rm rot}}\ll 1.
\label{p3p}
\end{equation}
Note that this model of strictly rectilinear flows amounts to ignoring nondissipative
contributions to the mutual friction force, ie it leads to $B'\equiv 0$.

\section{Dissipative vortex dynamics: phenomenological approach}\label{s3}
A microscopic view to the above lattice leads us to focusing on a single vortex dynamics.
Thus, we assume that the motion of each vortex is ruled by \cite{don}:
\begin{equation}
m_v \ddot{\bf r} = \rho_s\kappa\, {\bf \hat z}\times (\dot{\bf r}-{\bf v}_s)
-D(\dot{\bf r}-{\bf v}_n),\label{0.1}
\end{equation}
where ${\bf r}=x{\bf \hat x}+y{\bf \hat y}$ denotes the position of the vortex core
and $D$ denotes a microscopic friction 
coefficient arising from  vortex-quasiparticle scattering. 
The right-hand side of (\ref{0.1}) represents the total force acting on the vortex, namely
 the Magnus force (first term) plus 
the dissipative force (second term). Here we have disregarded
again any contribution from a nondissipative component of the mutual friction force\footnote{
Actually, according to some theories \cite{nonex}
 and related experimental evidence \cite{vin,zhu}, such a component may in fact be non-existent.}.
Now, replacing (\ref{1.1}), (\ref{1.2}), (\ref{1.6}) and $\Omega_s(t)\simeq\Omega_{\rm rot}$
 in (\ref{0.1})
 we obtain,
\begin{eqnarray}
\ddot x & = & -\Omega\frac{D}{\rho_s\kappa}\dot x-\Omega\,\dot y
-2\, \Omega_{\rm rot}\,\Omega\frac{D}{\rho_s\kappa} y
\left(1+\frac{w_0}{\Omega_{\rm rot}}e^{-\Omega_{\rm rot}Bt}\right) \label{0.2}\\
\ddot y & = & -\Omega\frac{D}{\rho_s\kappa}\dot y
+2\,\Omega\,\Omega_{\rm rot}y+\Omega\,\dot x.\label{0.3}
\end{eqnarray}
The equation (\ref{0.2}) can be formally solved in $\dot x(t)$,
\begin{eqnarray}
\fl\dot x(t)  =  e^{-(\Omega D/\rho_s\kappa)t}\,\dot x(0)
-\int_0^t d\tau 
e^{-(\Omega D/\rho_s\kappa)\tau}\nonumber \\
\times\left[\Omega\,\dot y(t-\tau)
+2\, \Omega_{\rm rot}\,\Omega\frac{D}{\rho_s\kappa} y(t-\tau)
\left(1+\frac{w_0}{\Omega_{\rm rot}}e^{-\Omega_{\rm rot}B(t-\tau)}\right)\right]
\label{0.4}
\end{eqnarray}
and replacing the above expression in (\ref{0.3}) we are led to a second-order integro-differential
 equation in $y(t)$. This problem, however, may be greatly simplified by utilizing the Markov approximation.
In fact, we first note that the coefficients in the arguments
of the exponentials in  (\ref{0.4}) define two very different
time scales, namely a microscopic one given by $(\Omega D/\rho_s\kappa)^{-1}$ and a macroscopic one given by
$(\Omega_{\rm rot}B)^{-1}$, ie
\begin{equation}
\frac{\rho_s}{\rho_n}
\frac{\Omega_{\rm rot}}{\Omega}\ll 1,
\label{0.6}
\end{equation}
where we have taken into account that the macroscopic and microscopic friction coefficients are 
related for  $T\lesssim 1$ K through  \cite{don},
\begin{equation}
B=\frac{2D}{\rho_n\kappa}.
\label{0.5}
\end{equation}
Note that the inequality (\ref{0.6}) is expected to break down at extremely low temperatures.
Now, assuming that $t$ in (\ref{0.4}) belongs to the macroscopic time scale,
we have that the exponential factor in the first term will be negligible, while the same exponential function,
which depends on $\tau$ in front of the integrand, will make that
only the lowest portion of the integration domain
($\tau\ll t$) gives a nonnegligible contribution to the integral. 
That is, we may safely approximate all the 
dependencies on $t-\tau$ in (\ref{0.4}) by $t$,
 and the upper limit of the integral by $+\infty$.
Finally, 
the Markov approximation to such an equation 
 reads as,
\begin{eqnarray}
\fl\dot x(t)  =  
-\int_0^\infty d\tau 
e^{-(\Omega D/\rho_s\kappa)\tau}
\left[\Omega\,\dot y(t)
+2\, \Omega_{\rm rot}\,\Omega\frac{D}{\rho_s\kappa} y(t)
\left(1+\frac{w_0}{\Omega_{\rm rot}}e^{-\Omega_{\rm rot}B\, t}\right)\right]\nonumber \\
=-\frac{\rho_s\kappa}{D}\dot y
-2\, \Omega_{\rm rot}\, y
\left(1+\frac{w_0}{\Omega_{\rm rot}}e^{-\Omega_{\rm rot}B\, t}\right),
\label{0.7}
\end{eqnarray}
which replaced in (\ref{0.3})  and using (\ref{0.5}) yields,
\begin{equation}
\frac{\rho_n B}{2\rho_s\Omega}\,\ddot y + \dot y +\frac{\rho_n}{\rho_s}
B\, w_0\, e^{-\Omega_{\rm rot}B\,t}\, y = 0.
\label{0.8}
\end{equation}
Here it is useful to change to the adimensional macroscopic time variable $\mathcal{T} =\Omega_{\rm rot}B\,t$,
\begin{equation}
\frac{1}{2}\left(\frac{\rho_n B}{\rho_s}\right)^2
\left(\frac{\rho_s}{\rho_n}
\frac{\Omega_{\rm rot}}{\Omega}\right)\frac{d\,^2y}{d\mathcal{T}^2}+ \frac{dy}{d\mathcal{T}}
+\left(\frac{\rho_n w_0}{\rho_s\Omega_{\rm rot}}\right) e^{-\mathcal{T}} y = 0
\label{0.9}
\end{equation}
and recall that $\rho_n/\rho_s\lesssim 10^{-3}$ and $B\lesssim 1$ for $T\lesssim 1$ K. Then, assuming
$w_0\lesssim \Omega_{\rm rot}$ and taking into account (\ref{p3p}) and (\ref{0.6}),
we realize that we may safely drop the term containing the second derivative in (\ref{0.9}), since 
we may estimate that the coefficient in front of such a derivative will be 
a quantity of third order in comparison 
 to the first-order small parameter in front of the exponential.
Note that according to (\ref{2}), this approximation turns out to be equivalent to the usual one of 
neglecting the vortex mass.
 Thus, we are led to a first-order 
differential equation which is easily integrated yielding,
\begin{equation}
y(t)= y(0)\,\exp[
(\rho_n w_0/\rho_s\Omega_{\rm rot})
(e^{-
\Omega_{\rm rot}Bt}-1)],
\label{0.4p}
\end{equation}
ie a small vortex displacement in the direction perpendicular to the velocity of the background superflow,
while (\ref{0.7}) becomes
\begin{equation}
\dot x = -2\Omega_{\rm rot} y,
\label{0.4pp}
\end{equation}
which simply states that the $x$-component of the vortex velocity will coincide with 
the superfluid velocity.

\section{Dissipative vortex dynamics: Hamiltonian approach}\label{s4}
\subsection{Vortex Hamiltonian}\label{sec1A}
We start from the vortex equation of motion (\ref{0.1}) in the absence of normal fluid
\begin{equation}
\ddot{\bf r}=\Omega\, {\bf \hat z}\times (\dot{\bf r}-{\bf v}_s),\label{1}
\end{equation}
which turns out to be analogous to that ruling the two-dimensional motion of a 
negative point charge 
in the presence of  magnetic and electric fields in the $z$ and $y$ directions, 
respectively \cite{yosh}.
Such an equation derives from the Hamiltonian
\begin{equation}
H_v = \frac{m_v}{2}(v_x^2+v_y^2)-\Omega_{\rm rot}\,\rho_s\kappa\, y^2\label{4},
\end{equation}
being (Landau gauge)
\begin{equation}
\eqalign{v_x = \frac{p_x}{m_v}-\Omega\, y \cr
v_y = \frac{p_y}{m_v},}\label{5} 
\end{equation}
where ${\bf p}=p_x{\bf \hat x}+p_y{\bf \hat y}$ corresponds to the vortex canonical momentum. 
Note that such a momentum as well as the Hamitonian (\ref{4}) are given per unit length of the vortex line.
Then, from 
Hamilton equations it is easy to check that the expressions (\ref{5}) correspond to the velocity of the
 vortex core
$\dot{\bf r}$, while the acceleration is indeed given by (\ref{1}). From the above 
electromagnetic analogy it is also useful to represent the coordinate ${\bf r}$ as the sum
of the center coordinate ${\bf r}_0=x_0{\bf \hat x}+y_0{\bf \hat y}$ 
of the cyclotron circle plus the relative coordinate 
${\bf r}'$
from such a center \cite{yosh}, being 
\begin{equation}
\eqalign{x_0 = -\frac{v_y}{\Omega}+x \cr
y_0 = \frac{v_x}{\Omega}+y.}\label{7} 
\end{equation}
Then, it is easy to extract the time evolution of such coordinates working to zero-th order in 
the small parameter $\Omega_{\rm rot}/\Omega$:
\begin{equation}
\dot{{\bf r}}_0  = -2\Omega_{\rm rot}y_0\,{\bf \hat x}
\label{p22a},
\end{equation}
which means that the center coordinate ${\bf r}_0$ will move with the superfluid velocity, while
${\bf r}'(t)$ will perform a 
counterclockwise 
circular cyclotron motion with angular frequency $\Omega$.
Finally, we may see from (\ref{7}) that the limit of a vanishing vortex mass
($\Omega\rightarrow \infty$), which is often found in
the literature \cite{fet1}, amounts to ignoring the cyclotron motion 
(${\bf r}'\rightarrow 0$, ${\bf r}\equiv {\bf r}_0$).

\subsection{Quasiparticle Hamiltonian}\label{sec1B}
Although we have described
 the vortex Hamiltonian in classical terms, 
 it is immediate by means of the electromagnetic analogy to switch to a quantum mechanical
picture  \cite{yosh}. This constitutes a
necessary generalization, since the
normal fluid will be treated as a low-temperature quantum gas. 
In fact, such a fluid will be represented by the following Hamiltonian:
\begin{equation}
H_n=\sum_{ {\bf q}}\,\hbar\omega_{ {\bf q}} \,\,
a_{{\bf q}}^\dagger \,  a_{ {\bf q}}
\label{19},
\end{equation}
where $a_{{\bf q}}^\dagger$ denotes a creation operator of quasiparticle excitations 
of (pseudo-) momentum $\hbar{\bf q}$ and frequency $\omega_{ {\bf q}}$. Measuring such a frequency
from the lab frame, it is written as a Doppler-shifted frequency from the superfluid frame,
$\omega_{ {\bf q}}=\omega_q+{\bf q}\cdot{\bf v}_s$, where $\omega_q$ is the familiar (isotropic)
dispersion relationship of $^4$He quasiparticle excitations. 
 Note also that we disregard any 
interaction between the 
quasiparticles themselves, since we
shall work at low enough temperature, so that they remain
dilute allowing their treatment as a noninteracting gas.

\subsection{Equation of motion in terms of pseudomomentum}\label{sec1C}
Note that according to the electromagnetic analogy, there are three kinds of momentum
to be ascribed to the vortex, viz the canonical one ${\bf p}$, the dynamical
one $m_v{\bf v}=\rho_s\kappa{\bf \hat z}\times{\bf r}'$ 
and the so-called {\em pseudomomentum} \cite{yosh}, which is given by
${\bf K}=-\rho_s\kappa{\bf \hat z}\times{\bf r}_0$ and should be regarded as the generator of
translation. 
Then, adding
such a pseudomomentum to the quasiparticle pseudomomentum $\sum_{ {\bf q}}\,\hbar{\bf q} \,\,
a_{{\bf q}}^\dagger \,  a_{ {\bf q}}$, we have the pseudomomentum of the whole system.
Note that only the $x$-component of the vortex pseudomomentum
$\rho_s\kappa\,y_0$
will commute with the vortex Hamiltonian (\ref{4}), unless $\Omega_{\rm rot}=0$. 
This result may be easily interpreted, since
a superflow of velocity ${\bf v}_s=-2\,\Omega_{\rm rot}\,y\,{\bf \hat x}$ 
produces a translation symmetry breaking in
the $y$-direction.

Next let us analize 
the conservation
 of the $x$-component of the pseudomomentum for a given vortex of the lattice
and its surrounding quasiparticles\footnote{Here it is worth comparing with
a recently derived conservation theorem for wave pseudo\-momentum
plus a suitably defined classical vortex impulse  \cite{b-m}.}:
\begin{equation}
\rho_s\kappa L\langle y_0\rangle+
\sum_{ {\bf q}}n_{{\bf q}}\,\hbar{\bf q}\cdot{\bf \hat x} = {\rm const.},\label{3.1}
\end{equation}
where $n_{\bf q}=\langle a_{{\bf q}}^\dagger \,  a_{ {\bf q}}\rangle$ denotes
the average number of quasiparticles with pseudomomentum $\hbar{\bf q}$ and
$L$ denotes the vortex line length.
We shall assume that such a population is well described by a local equilibrium form:
\begin{eqnarray}
n_{\bf q}&=&[e^{\hbar\omega_{\bf q}/k_BT}-1]^{-1}
\simeq[e^{\hbar\omega_q/k_BT}-1]^{-1}\nonumber\\
&+&
\frac{\hbar\,{\bf q}\cdot[{\bf v}_n-{\bf v}_s]}{4k_BT\sinh^2(\hbar\omega_q/2k_BT)},
\label{31}
\end{eqnarray}
where the quasiparticle frequency is measured from a reference frame where 
the local normal fluid is at rest,
$\omega_{\bf q}=\omega_q-{\bf q}\cdot[{\bf v}_n-{\bf v}_s]$, with
${\bf v}_s=-2\,\Omega_{\rm rot}\,y\,{\bf \hat x}$ and
${\bf v}_n=-2\,\Omega_n(t)\,y\,{\bf \hat x}$, [$\Omega_n(t)$ given by (\ref{1.6})].
Then, replacing (\ref{31}) in (\ref{3.1}) we have,
\begin{equation}
\rho_s\kappa\langle y_0\rangle+
A_v\,\rho_n
({\bf v}_n-{\bf v}_s)\cdot {\bf \hat x}= {\rm const.},\label{3.3}
\end{equation}
where 
\begin{equation}
\rho_n=\frac{\hbar^2}{12A_v\, L\,k_BT}\sum_{ {\bf q}}\,\frac{q^2}{\sinh^{2}(\hbar\omega_q/2k_BT)}
\label{3.4}
\end{equation}
yields the normal fluid density and $A_v=(2\Omega_{\rm rot}/\kappa)^{-1}$ corresponds to the area
per vortex of the lattice (Feynman's rule). 
Taking into account that the velocity fields in (\ref{3.3}) must be evaluated
at $y=\langle y_0\rangle$, and using the solutions (\ref{1.6}) and (\ref{1.7}) we 
obtain,
\begin{equation}
\langle y_0(t)\rangle=\langle y_0(0)\rangle\,
[1+(\rho_n w_0/\rho_s\Omega_{\rm rot})
(e^{-
\Omega_{\rm rot}Bt}-1)],
\label{3.4p}
\end{equation}
which according to  (\ref{p3p}), 
turns out to be equivalent to (\ref{0.4p}). In other words, starting from the
conservation of pseudomomentum for a vortex and its surrounding quasiparticles, without any further assumptions
regarding the vortex mass,
we have shown that the $y$-component of ${\bf r}_0$ presents 
the same dissipative evolution as that obtained in section \ref{s3}
for a  vortex of negligible mass, for which ${\bf r}\equiv {\bf r}_0$.
Thus, we may see how the dynamics in terms of the pseudomomentum makes the coordinate
${\bf r}_0$ the focus of attention, while the less relevant cyclotron dynamics of ${\bf r}'$,
which was forced to vanish in the phenomenological treatment
through the neglect of the vortex mass,
is now naturally ignored. 

To study the dynamics along the $x$-direction we must focus on
the $y$-component of pseudomomentum, which is given by the nonconserved vortex 
contribution $-\rho_s\kappa\langle x_0\rangle$ (cf (\ref{p22a})). 
Finally, taking into account
the whole vector
pseudomomentum ${\bf K}_{\rm tot}$ of the vortex plus 
surrounding quasiparticles,
we may write
the following equation:
\begin{equation}
\frac{d{\bf K}_{\rm tot}}{dt}={\bf F}_{\rm ext},\label{ps1}
\end{equation}
where
\begin{equation}
{\bf K}_{\rm tot}=-\rho_s\kappa{\bf \hat z}\times\langle{\bf r}_0\rangle+
\frac{\kappa}{2\Omega_{\rm rot}}\,\rho_n
({\bf v}_n-{\bf v}_s)
\end{equation}
and 
\begin{equation}
{\bf F}_{\rm ext}=-\rho_s\kappa{\bf \hat z}\times{\bf v}_s
\end{equation}
represents the force exerted on the vortex by the superfluid current \cite{sonin},
which is responsible for the nonconserved $y$-component of pseudomomentum.

The equation of motion (\ref{ps1}) constitutes the central result of this paper, which
may be regarded as arising from a straightforward combination of the electromagnetic analogy 
and the
fruitful concept of pseudomomentum. 
Here it is important to remark that contrary to the phenomenological equation of motion
(\ref{0.1}), the pseudo\-momentum equation (\ref{ps1}) does not involve 
the vortex mass, a fact to be regarded as most welcome, given its 
apparent elusive and ambiguous nature \cite{thou}.

To conclude we may observe that
it would be important to generalize this picture 
for the rotating superfluid, where the key magnitude would be represented 
by an {\it angular 
pseudomomentum}. To achieve this goal, given the drawbacks of the 
HVBK equations pointed out in section \ref{s1}, it would be helpful previously to seek for a validation of
the present results without utilizing such equations as a starting point.

\ack
This work has been performed under Grant PIP 5409 from CONICET, Argentina.

\end{document}